\documentclass[preprint,12pt]{aastex}

\shortauthors{Hirano and Gwinn}
\shorttitle{A Ray-Tracing Model of the Vela Pulsar}

\def\bbeta{\mbox{\boldmath$\beta$}}

\def\th0{\theta_0}
\def\gp{\gamma_p}
\def\kpar{k_\shortparallel}
\def\kprp{k_\perp}
\def\ompp{\omega'_p}
\def\sigt{\sigma_\theta}

\begin{document}

\title{A Ray-Tracing Model of the Vela Pulsar}

\author{C. Hirano and C. R. Gwinn}

\affil{Department of Physics,
       University of California,
       Santa Barbara, CA 93106}
\email{placebo@condor.physics.ucsb.edu}
\email{cgwinn@condor.physics.ucsb.edu}

\begin{abstract}
In the relativistic plasma surrounding a pulsar, a subluminal
ordinary-mode electromagnetic wave will propagate along a magnetic
field line.  After some distance, it can break free of the field line
and escape the magnetosphere to reach an observer.  We describe a
simple model of pulsar radio emission based on this scenario and find
that applying this model to the case of the Vela pulsar reproduces
qualitative characteristics of the observed Vela pulse profile.
\end{abstract}


\clearpage

\section{Introduction}

Soon after the discovery of pulsars, \citet{Gol69} deduced the basic
structure of the magnetosphere.  The strong corotating magnetic field
of order $10^8-10^{12}~\rm{G}$ dominates the physics in the
surrounding region.  The rotating magnetic field produces forces
sufficient to rip electrons from the stellar surface with high energy.
The high-energy electrons emit photons which in turn quickly decay via
pair production.  This process repeats until the energies fall
sufficiently to halt pair production.  The consequent corotating
charge distribution cancels the ${\bf v}\times{\bf B}$ forces, except
perhaps at certain gaps, and the intense magnetic field constrains the
electrons to move along the field lines.  The end result is a
one-dimensional relativistic electron-positron wind flowing out along
the corotating magnetic field lines.  Those particles on the field
lines which extend beyond the light cylinder can escape to infinity.

Connecting this theoretical picture with observation has proven to be
difficult; we still do not understand the physical mechanism which
generates pulsar radio emissions.  Observations of pulsars reveal that
the individual pulses from a pulsar vary greatly from pulse to pulse
with no apparent pattern.  Averaging several hundred pulses, however,
yields a stable profile characteristic of a pulsar, yet these
characteristic profiles still differ widely from pulsar to pulsar
\citep{Lyn98}.  \citet{Ran83} has proposed that pulsars exhibit two
distinct types of emissions, core and conal, suggesting that more than
one emission mechanism may be at work.

Despite the large variety of integrated pulse profiles, the profiles
commonly exhibit a high degree of linear polarization.  The
polarization position angle often sweeps smoothly through a range of
angles of up to 180\arcdeg \citep{Lyn98}.  In their rotating-vector
model, \citet{Rad69} hypothesized that the radiation was polarized
along or orthogonal to the magnetic field lines; the position angle
therefore reflects the geometry of a dipolar magnetic field.

While some pulsars exhibit an abrupt jump between orthogonal modes of
polarization, their position angle distributions indicate that the
emissions contain two orthogonal components, each of which appear
consistent with the rotating-vector model \citep{Man75,Bac76}.
Fluctuations in the flux densities of the two modes would be observed
as a sudden jump in position angle \citep{McK98}.

The pulse also typically broadens with decreasing frequency, a
behavior usually attributed to radius-to-frequency mapping:
lower-frequency emissions are thought to originate farther away from
the stellar surface than higher-frequency emissions (see \citet{Cor78}
and references therein).  Because the magnetic field lines diverge,
the higher altitude translates to an increase in the size of the
emission region which in turn translates to a wider observed pulse at
lower frequency.

\citet{Bar86} investigated the propagation of electromagnetic waves in
the relativistic plasma permeating the pulsar magnetosphere and found
that one mode, the subluminal O-mode, follows paths along the
magnetic field lines.  A possible source of this
radiation is the pair-production front, where the primary particles
decay into those which make up the outflowing plasma
\citep{Ler70,Mel96}.  \citet{Lyu98} has also suggested that cyclotron
instabilities in the plasma will produce subluminal O-mode radiation
in the magnetosphere of a pulsar.

Observations of rays originating in this mode would reflect the
geometry of the dipolar field \citep{Gal96}, which could account for
some observed pulse shapes and position angle sweeps.  In this paper,
we consider a model that assumes a neutron star with a corotating
dipolar magnetic field and a source of radio-frequency radiation near
the stellar surface.  By determining the path of rays from the stellar
surface to the observer, we attempt to reproduce the pulse intensity
and polarization profiles of a pulsar, specifically the Vela pulsar.

\section{Wave Propagation in the Pulsar Magnetosphere}

In a relativistic plasma and a strong magnetic field ${\bf B}$,
electromagnetic waves propagate in two independent, linearly polarized
modes.  The electric field of an ordinary mode or O-mode wave lies in
the $({\bf k},{\bf B})$-plane, where ${\bf k}$ is the wave vector of
the wave; the electric field of the extraordinary mode or X-mode
points along the normal to this plane.  Because the charges can move
only along the field lines, the plasma has no effect on the X-mode, so
this mode propagates as if in a vacuum.

\begin{figure}
\plotone{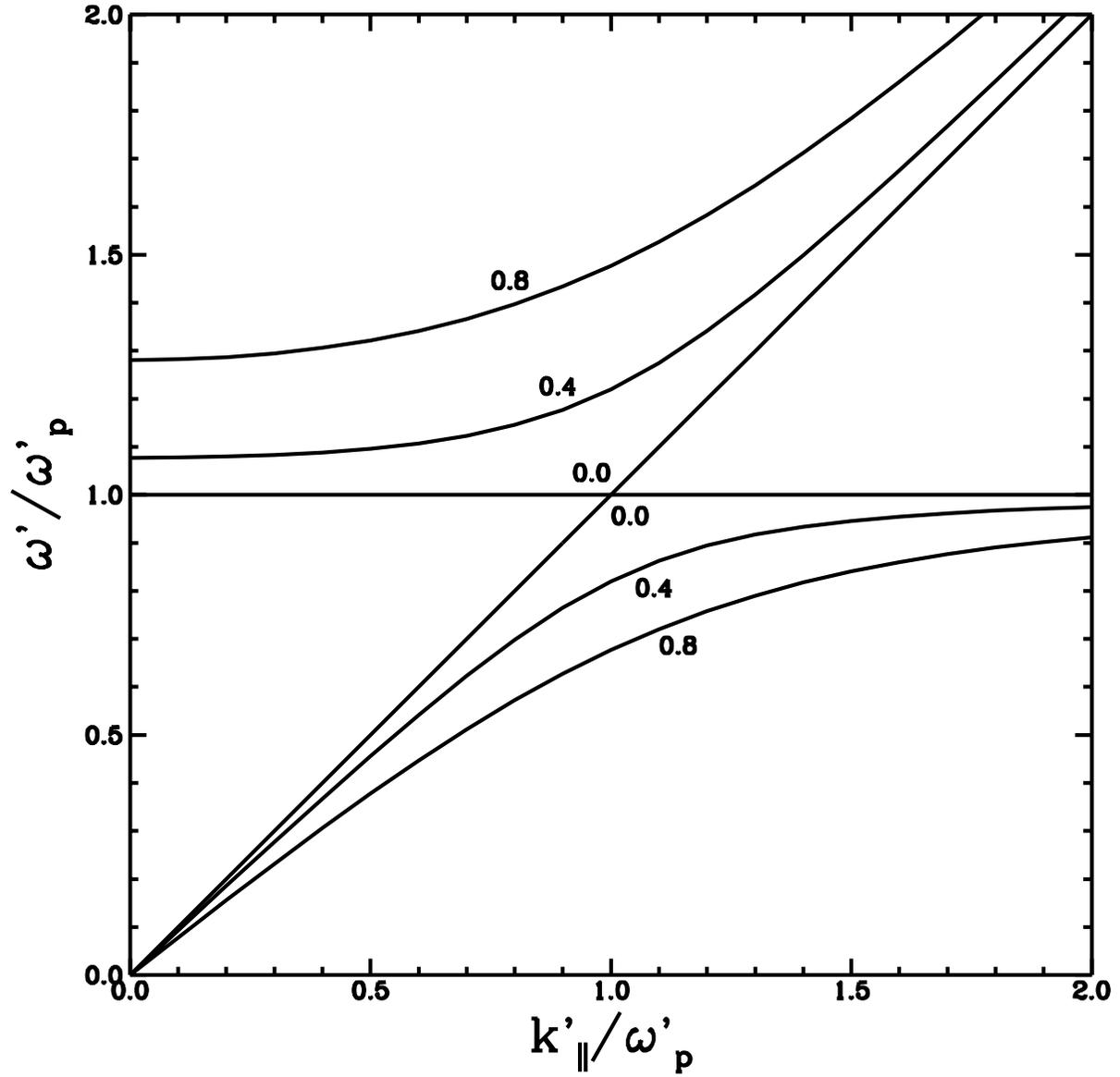}
\caption{Plot of the cold plasma O-mode dispersion relation in the rest frame
of the plasma for various values of $\kprp'/\ompp$, as indicated by the numbers
labeling each curve. The upper branch corresponds to the superluminal or fast
mode; the lower branch, to the subluminal or slow mode.}
\label{disp_rel}
\end{figure}

In the limit of an infinitely strong magnetic field, the dispersion
relation for an O-mode wave of frequency $\omega$ propagating in a
cold plasma with plasma frequency $\omega_p$ and consisting of
electrons and positrons moving with speed $\beta_p$ is given by
\begin{equation}
(\omega^2-\kpar^2)\left(1-\frac{\omega_p^2}{\gp^3\omega^2[1-\beta_p(\kpar/\omega)]^2}\right)-k^2_\perp
= 0
\end{equation}
where $\gp$ is the usual relativistic factor and $\kprp$ and $\kpar$
are respectively the components of the wave vector normal to and along
the magnetic field \citep{Aro86}.  Figure~\ref{disp_rel} shows a plot
of the dispersion relation in the rest frame of the plasma for various
values of $\kprp'/\ompp$.  Here the primes denote quantities measured
in the rest frame of the plasma.  For each value of $\kprp'/\ompp$,
the figure shows two curves; the upper branch corresponds to the
superluminal or fast mode while the lower branch corresponds to the
subluminal or slow mode.

Each ray begins its journey as a slow-mode wave and either transitions
to the fast mode and escapes the magnetosphere or succumbs to Landau
damping and dissipates in the plasma.  We assume the transition occurs
via some mechanism, when it reaches a point such that the distance
between a pair of curves is minimized, or in other words, when the two
types of waves are most alike. We approximate this criterion by
requiring the transition to occur when $\kpar'/\ompp=1$.

Several points should be made here.  First, the criterion which
determines where the ray escapes from the magnetosphere will obviously
affect the observed pulse shape. The purpose of this paper, however,
is not to explore the physics of the mode conversion but rather to
examine how refractive effects coupled with the geometry of the dipole
field might give rise to features of the pulse profile.  Second, the
transition criterion we chose corresponds to mode conversion in the
weak-turbulence limit \citep{Lyu98}.  Turbulence, in effect, smears
out the dispersion curves, and the transition occurs via the region in
which the smeared curves overlap.  Third, the conversion rate may be
so low in absolute terms so that it could not account for the observed
radio emission. Clearly, the physics of the mode conversion is
critical for a complete understanding pulsar radio emission and has
been investigated (see, for example, \citet{Bes88}).  However, such a
discussion is beyond the intended scope of this paper.

We must also consider that the subluminal mode may interact with the
electrons and positrons and transfer its energy to the plasma.  The
mode conversion can not occur if Landau damping dissipates the wave
before it can reach its would-be transition point, but \citet{Aro86}
found that for canonical pulsar parameters, Landau damping would not
be significant if the mode conversion occurred at radii less than 1000
km.  We find that the transition points lie at altitudes significantly
less than 1000 km and therefore neglect Landau damping.

In terms of ${\ompp}^2 = \omega_p^2/\gp$ and $\omega' =
\gp(\omega-\beta_p\kpar)$,
where again the primes denote quantities measured in the rest frame of
the plasma, the dispersion relation becomes
\begin{equation}
(\omega^2-\kpar^2)\left(1-\frac{{\ompp}^2}{\omega'^2}\right)-k^2_\perp = 0.
\end{equation}
It follows that for the subluminal mode because $\omega'<\ompp$, we
must have $|\omega|<|\kpar|$.  In the corotating frame, the plasma is
essentially
moving at the speed of light and outruns the wave; boosting to the rest
frame of the plasma therefore yields a negative frequency.  Consequently,
the relevant branch of the dispersion relation is given by
\begin{equation}
\omega' =
-\sqrt{\frac{{\ompp}^2+\kpar'^2+\kprp'^2-\sqrt{({\ompp}^2+\kpar'^2+\kprp'^2)^2-4{\ompp}^2\kpar^2}}{2}}
\label{disp_eq}
\end{equation}
To determine how the wave evolves, we must know how $\kprp'$ and
$\ompp$ change as the wave travels away from the stellar surface.

\begin{figure}
\plotone{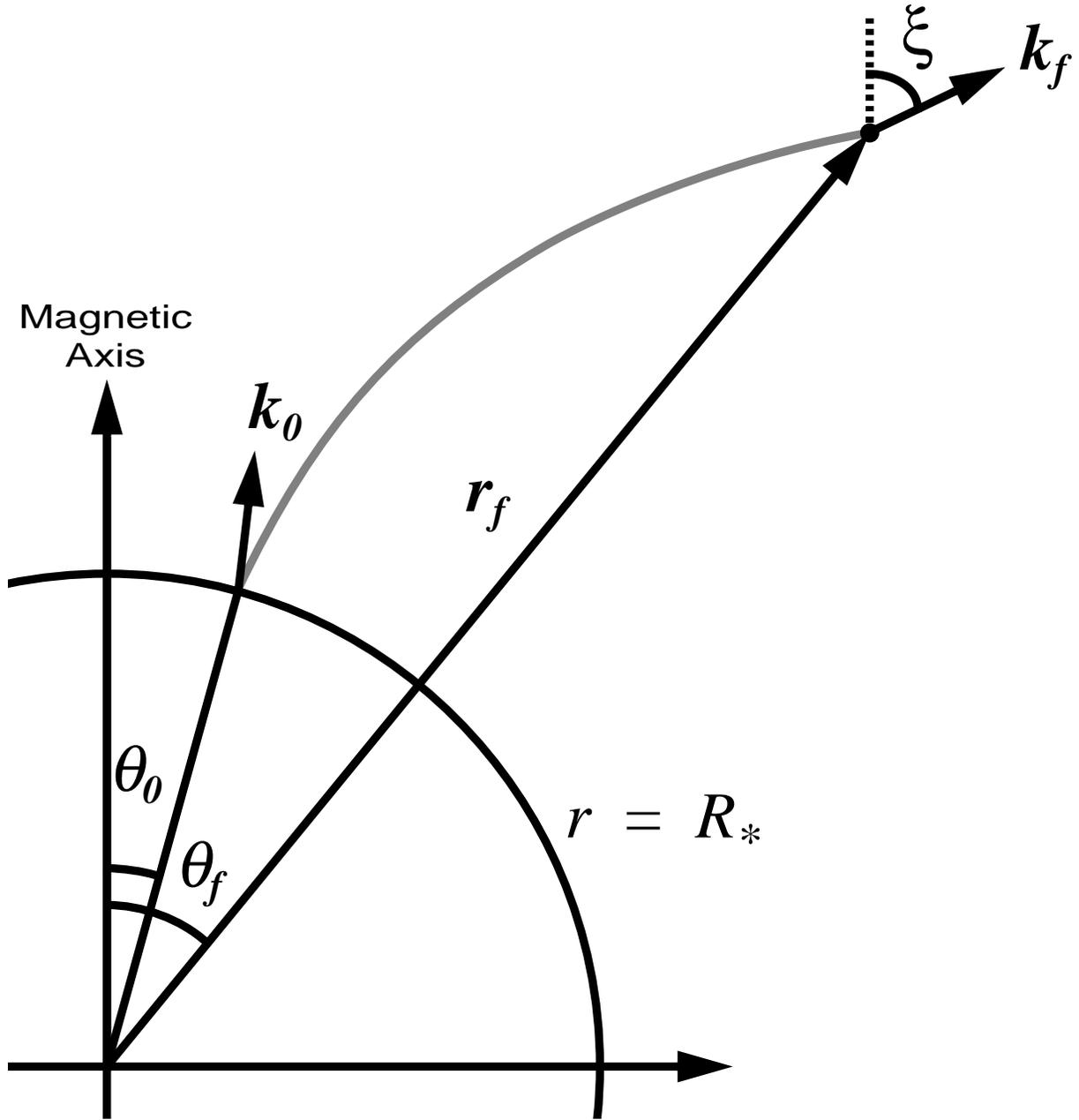}
\caption{As shown by the halftone line, the slow-mode wave leaves the
stellar surface at $(R_\ast,\th0)$ and follows a magnetic field line.
At $(r_f,\theta_f)$, it transitions to the fast mode and travels along
$\bf k_f$.}
\label{mag_geom}
\end{figure}

The slow O-mode wave packets obey the relations \citep{Bar86}:
\begin{mathletters}
\begin{eqnarray}
\frac{r}{r_0} & = & \left(\frac{\theta}{\th0}\right)^2 \label{dpf} \\
n_\perp \equiv \frac{\kprp}{\omega} & =
&\frac{3\th0}{8}\left(\frac{r}{r_0}\right)^{1/2}-\left(\frac{3\th0}{8}-n_{\perp_0}\right)\left(\frac{r}{r_0}\right)^{-3/2}
\end{eqnarray}
\end{mathletters}
where $r$ and $\theta$ are the spherical coordinates of the wave
packet in the corotating frame, with the dipole axis pointing along
the $z$-axis, as illustrated in Figure~\ref{mag_geom}.  Here, the
0~subscript denotes values at the stellar surface.  The first equation
simply says that the wave packet follows the magnetic field lines.
The second equation describes how the direction of the wave vector
relative to the field line evolves as the wave packet moves down the
field line.  As the wave moves farther away from the neutron star, the
orientation of the wave vector depends less on its initial orientation
and increasingly on its displacement from the star.

The strong magnetic field constrains the electrons and positrons to
move along the field lines; hence, their density is proportional to
the magnetic field strength which, for a dipolar magnetic field,
varies as $r^{-3}$.  In the corotating frame then, the plasma
frequency is given by
\begin{equation}
\omega_p^2(r) = 2\frac{4\pi N_0 e^2}{m}\,\left(\frac{r_0}{r}\right)^3
\end{equation}
where $N_0$ is the plasma density at $r=r_0$, $e$ is the elementary
charge, and $m$ is the mass of the electron.  The factor of two
reflects the fact that the plasma consists of positrons and electrons.
Transforming to the comoving frame introduces a factor of $1/\gp$
because of length contraction, giving
\begin{equation}
\omega'^2_p(r) = \frac{8 \pi N_0 e^2}{\gp m}\,\left(\frac{r_0}{r}\right)^3 =
{\omega'_{p_0}}^2 \left(\frac{r_0}{r}\right)^3
\end{equation}
where $\omega'_{p_0}=\ompp(r_0)$.  In reality, the magnetosphere may
contain regions where $\omega_p$ does not scale in this manner, and
refraction may occur into these regions.  Our model neglects this
possibility.

\begin{figure}
\plotone{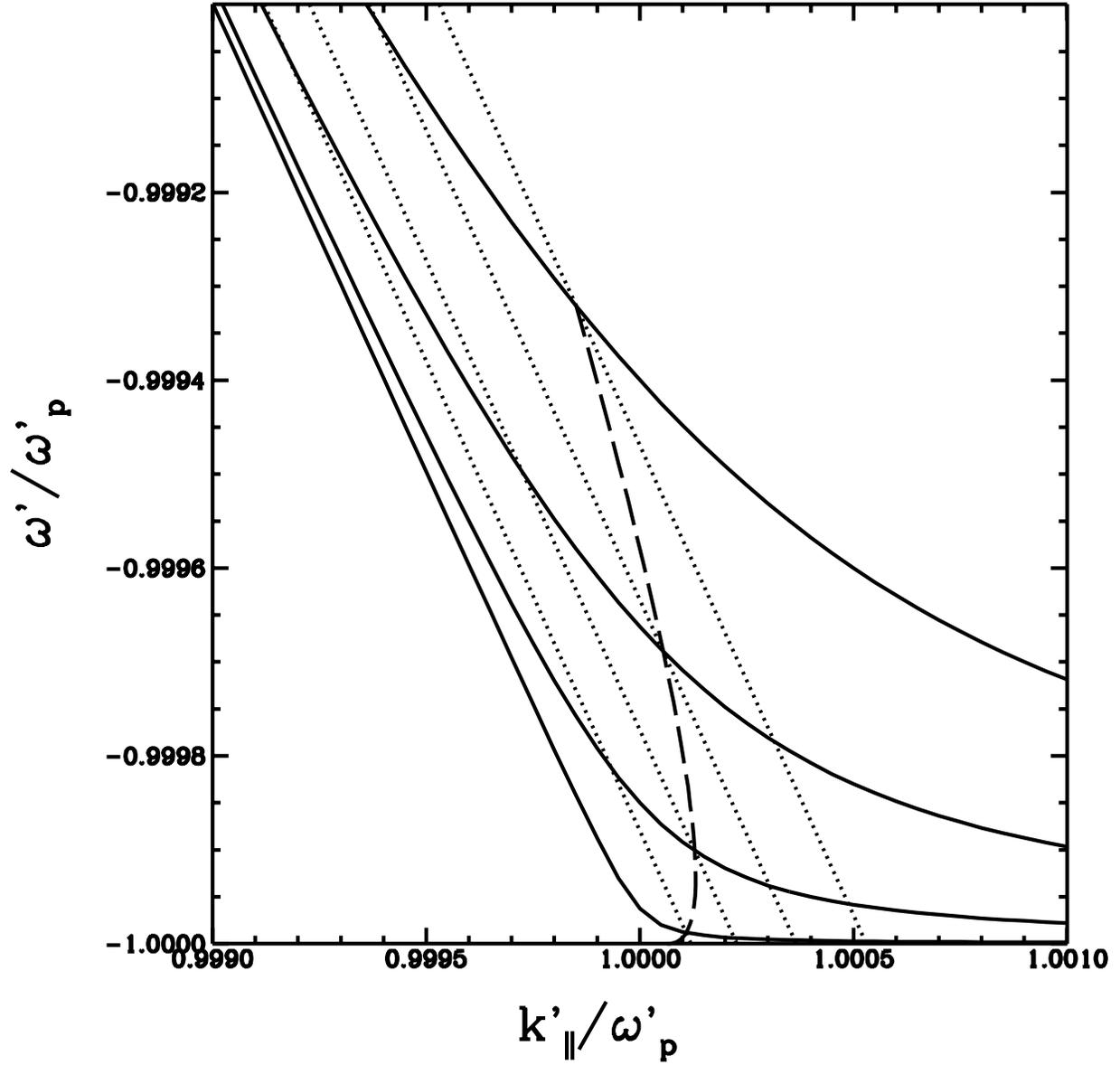}
\caption{A line and a dispersion relation curve are associated with
each point in the trajectory of an O-mode wave packet.  The
intersection of the two yields the waves values of $\omega'$ and
$\kpar'$ at that point.  The dashed line shows how $\omega'$ and
$\kpar'$ evolve as the wave propagates away from the pulsar.}
\label{evolution}
\end{figure}

Knowledge of both $\kprp'$ and $\ompp$ at each point on the trajectory
allows us to determine which dispersion relation curve the wave
satisfies but not the specific values of $\omega'$ and $\kpar'$.  To
find these quantities, we note that the wave also satisfies the
Lorentz transformation
\begin{equation}
\omega = \gp(\omega'+\beta_p\kpar')
\label{lorentz_eq}
\end{equation}
where $\omega$ is constant in the geometric-optics limit.  Plotted on
the same axes as the dispersion relation, this relation corresponds to
a line with slope $-\beta_p$ and intercept $\omega/\gp\ompp$.  The
intersection of this line with the dispersion relation curve gives us
the values of $\omega'$ and $\kpar'$ for the slow-mode wave at a given
radius.  Figure~\ref{evolution} shows four pairs of curves,
corresponding to distances of ten, twenty, thiry, and forty stellar
radii, for a ray following a field line.  The dashed curve shows how
$\omega'$ and $\kpar'$ evolve as the ray travels away from the stellar
surface.

To find the transition point, we expand equation~(\ref{disp_eq}) in
powers of $\kprp'/\ompp$ with $\kpar'/\ompp=1$ to find
\begin{equation}
\frac{\omega'}{\ompp} \approx -1+\frac{1}{2}\frac{\kprp'}{\ompp} =
-1+\frac{3}{16}\frac{\omega}{\omega'_{p_0}}\th0\left(\frac{r}{r_0}\right)^2
\end{equation}
Similarly, from equation~(\ref{lorentz_eq}), we have
\begin{equation}
\frac{\omega'}{\ompp} = -\beta+\frac{\omega}{\gp\ompp}
=-\beta+\frac{\omega}{\gp{\omega'_{p_0}}}\left(\frac{r}{r_0}\right)^{3/2}
\end{equation}
Setting these two expressions equal to each other gives
\begin{equation}
1-\beta+\frac{\omega}{\gp{\omega'_{p_0}}}\left(\frac{r}{r_0}\right)^{3/2}-\frac{3}{16}\frac{\omega}{\omega'_{p_0}}\th0\left(\frac{r}{r_0}\right)^2=0
\label{rho}
\end{equation}
which may be solved numerically for $r/{r_0}$.

After the mode conversion, a ray escapes the magnetosphere as a
vacuum electromagnetic wave.  The angle between the wave vector and
the field line is approximately $n_\perp \approx (3/8)\theta$, and the
angle between the field line and the dipole axis is $(3/2)\theta$.
Therefore, after the transition occurs, the wave travels along a line
that makes an angle $\xi = (9/8)\theta_f$ with the axis of the dipolar
field, where $\theta_f$ is the value of $\theta$ at the mode
transition.

We simplistically assume in our model that the direction of the wave vector
and the polarization remain unchanged by the mode conversion.  If
the two modes couple due to turbulence within the
plasma, one would not expect such a conversion
mechanism to preserve ${\bf k}$ and the polarization.  Clearly, even
in this simple model, the direction of the wave vector must change
during the transition; because the phase velocities of the two modes
are different, the constraints on $\omega$ and $\kprp$ require that
the two modes must have different $\kpar$s.  The exact manner in which
the slow mode maps to the fast mode depends on details of the
conversion process, a discussion of which is beyond the scope of this
paper.

\section{Ray Trajectories}

\begin{figure}
\plotone{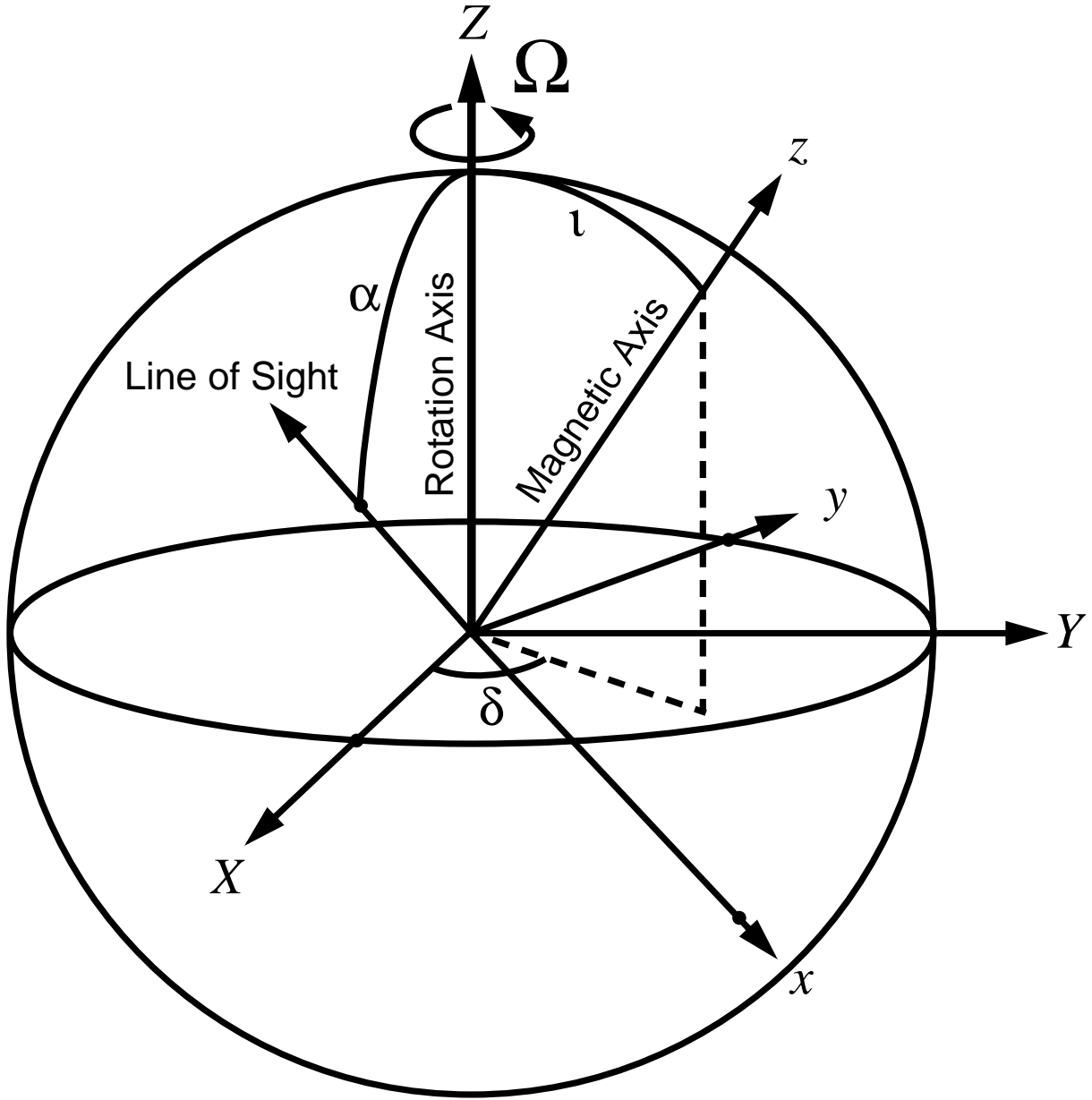}
\caption{The axis of rotation points along the $Z$-axis.  The line of
sight lies in the $X,Z$-plane and makes an angle $\alpha$ with the
rotation axis, and the direction of the magnetic axis is given by the
polar angle $\iota$ and the azimuthal angle $\delta=\Omega t$.}
\label{obs_geom}
\end{figure}

To analyze the path of a wave in the magnetosphere, it is convenient
to focus on two reference frames: the observer's frame and the
corotating frame.  Figure~\ref{obs_geom} illustrates the relationship
between cartesian coordinates in the two frames.  The set denoted by
capital letters corresponds to the observer's frame.  The $Z$-axis
points along the rotation axis of the pulsar, and the line of sight
lies in the $XZ$-plane.  The second set, denoted by lower-case
variables, lives in the corotating frame.  It is related to the first
set by a rotation by $\iota$ about the $Y$-axis followed by a rotation
by $\delta$ about the $Z$-axis, where the angles $\delta$ and $\iota$
describe the orientation of the pulsar.  In this system, the magnetic
axis coincides with $z$-axis, and the rotation axis lies in the
$xz$-plane.  Finally, because we are working with a dipolar field, the
corresponding set of spherical coordinates $(r,\theta,\phi)$ in the
corotating frame is useful.

The transition from slow to fast mode may occur at a distance which is
a significant fraction of the light cylinder radius; consequently, we
must take into account the effects of the boost from the comoving
frame to the observer's frame on the wave vector of the ray.  The
boost changes both the frequency and the direction of the ray.  The
frequency shift is small, about 1\%, and is relatively unimportant.
The change in direction, on the other hand, is more significant since
the final direction of the ray in the observer's frame determines if
and when the ray will reach the observer.

Each point on the transition surface appears to emit a beam of light
with a fixed direction in the corotating frame.  In the observer's
frame, the beams rotate with the star.  If a beam makes an angle
$\alpha$ with the rotation axis, it will sweep across the line of
sight when the star has rotated to an appropriate orientation, and the
beam will reach the observer.

\begin{figure}
\plotone{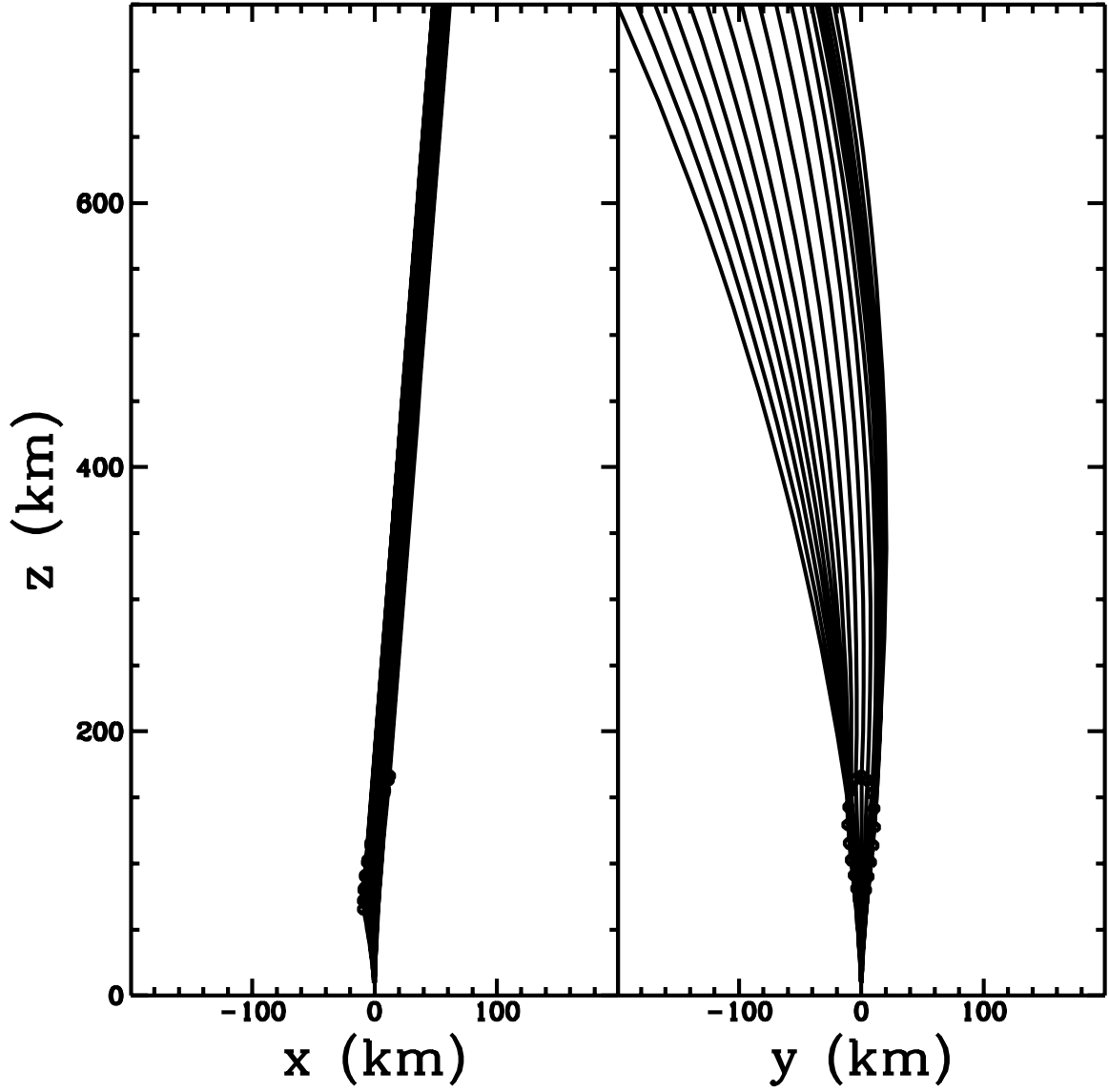}
\caption{These figures show the paths of several rays, as seen in the
corotating frame, projected onto the $xz$- and $yz$-planes.}
\label{raytrace}
\end{figure}

Figure~\ref{raytrace} illustrates the paths taken by several of these
rays as seen from the corotating frame.  Up to the transition point,
the rays follow the magnetic field lines; after the transition, they
travel in the directions given by their wave vectors.  In the
observer's frame, the rays will travel in straight lines after the
transition.  In the corotating frame, they will follow spiral
trajectories.

\section{Constructing the Pulse Profile}

To construct the pulse profile, we start with a uniform distribution
of rays on the stellar surface centered on the magnetic pole, trace
their paths, and select only the ones which will reach the observer.
The pulse period is divided into some number of equally-sized bins,
and the time of arrival of a ray determines into which bin it goes.
Averaging the position angles of the rays contributing to a bin yields
the final position angle for that bin.

Each ray carries a Gaussian weighting, with the distribution centered
on $\theta_0=0$.  The width $\sigt$ of the Gaussian corresponds
physically to the size of the beam rising from the stellar surface.
Appealing to the central limit theorem, we used a Gaussian distribution
so that the resulting profile is not that of a single pulse but that of
many averaged together.  In other words, the model produces the
integrated pulse profile.

\begin{figure}
\plotone{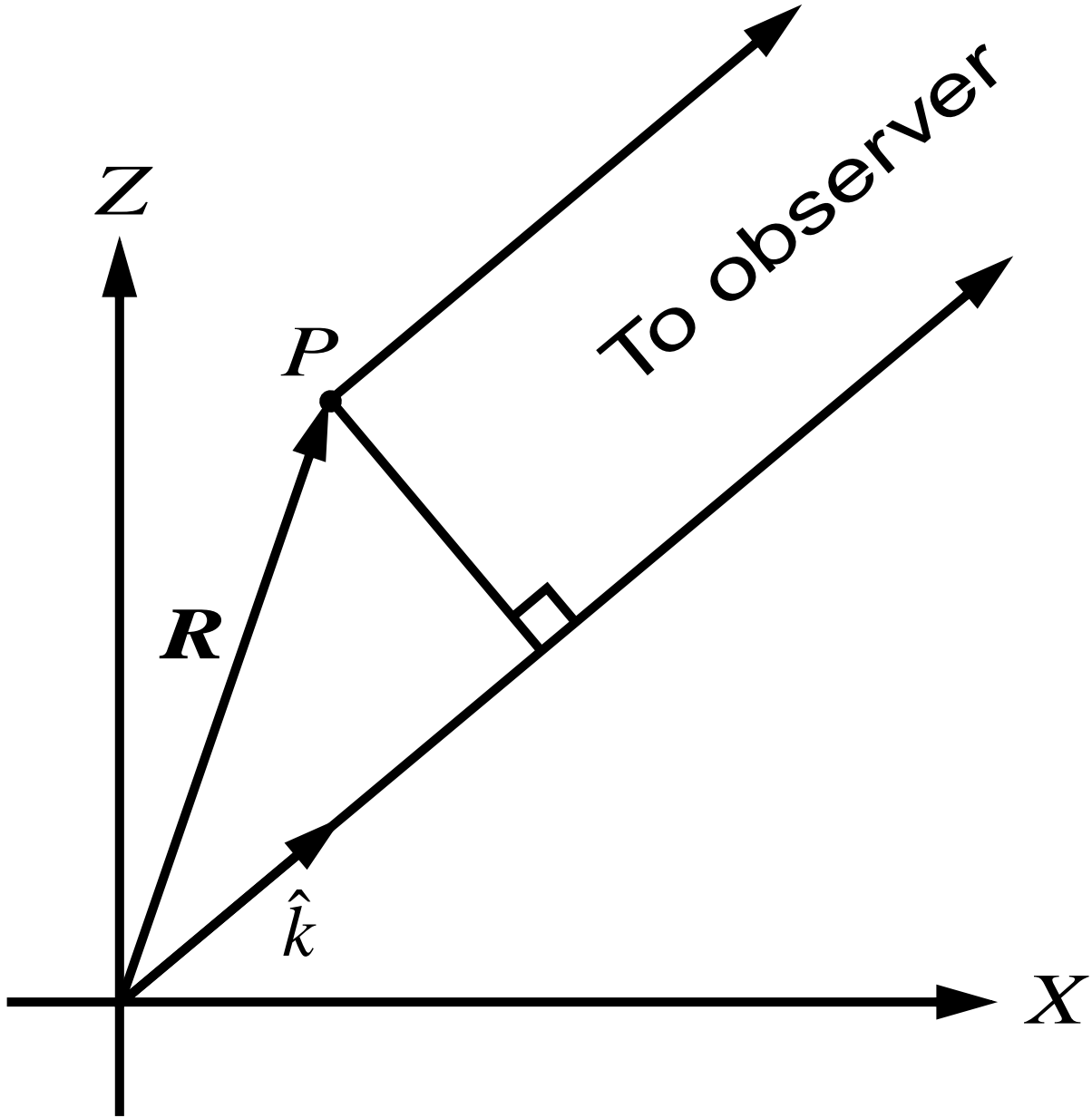}
\caption{The ray leaving from point \textsl{P} travels a distance
${\bf R}\cdot\hat{\bf k}$ less than a ray from the origin.  Hence,
it will arrive at the observer a time
$\delta t={\bf R}\cdot\hat{\bf k}/c$ earlier.
\label{tofa}}
\end{figure}

For a particular orientation of the neutron star, namely $\delta=0$,
we calculate for each ray its wave vector $\bf K$ in the frame of the observer.
The time of arrival for a ray that will reach the observer is then given by
\begin{equation}
t_{arr} = -\frac{\phi_K}{\Omega} - \frac{{\bf R}\cdot\hat{\bf K}}{c}
\end{equation}
where $\phi_K$ is the azimuthal angle of the wave vector,
$\Omega$ is the rotational speed of the pulsar, $\bf R$
is the point where the ray breaks free of the magnetic field line, and
$\hat{\bf K}$ is the unit vector in the direction of the wave vector.
The first term accounts for the fact that the star must rotate to the proper orientation for the ray
to align with the line of sight.  The second term arises from
the displacement of the transition
point from the star, as illustrated in Figure~\ref{tofa}.  The travel
time for the light ray from a pulsar at a distance $D$ is
approximately $t = D/c$; the actual travel time differs from this
value because the wave decouples from the plasma at a displacement
$\bf R$ from the center of the star.

As mentioned earlier, the rotating-vector model plausibly accounts for
the characteristic sweep in position angle under the assumption that
the polarization reflects the geometry of the dipolar field.  The
refraction model of wave propagation naturally explains the alignment
of the electric field with the field lines because, for both the
O-mode and X-mode, the direction of the electric field is determined
by the directions of the magnetic field and the wave vector $\bf k$.

For a dipolar field, the rotating-vector model predicts that the
position angle $\psi$ will vary according to
\begin{equation}
\tan\,\psi(\delta)=\frac{\sin\alpha\sin\delta}{\sin\iota\cos\alpha-\cos\iota\sin\alpha\cos\delta}
\end{equation}
where $\delta=\Omega t$.  This relation, however, neglects
relativistic and time-of-arrival effects.  The transformation from the
corotating frame to the observer's frame changes the direction of
$\bf E$.  Moreover, because of the second contribution to the
time-of-arrival delay, a ray corresponding to an orientation
$\delta_1$ can arrive before, at the same time as, or after a ray
corresponding to $\delta_2>\delta_1$. These effects may help explain
the deviations observed by \citet{Kri83} from the predictions of the
basic rotating-vector model.

We obtain the polarization of a ray in the observer frame by
calculating the electromagnetic field tensor $F_{\mu\nu}$ for the wave
in the corotating frame and then transforming it to the observer
frame.  To find the electric and magnetic field components in the
corotating frame, we define a set of cartesian coordinate axes so that
the $z$-axis points along the wave vector and the $x$-axis lies in the
$({\bf k},{\bf B})$-plane.  The electric field then points along rotated
$x$-axis, and the magnetic field, along the $y$-axis.

We have assumed that the electric field of a ray lies in the plane of
the dipole field line.  \citet{Che79} found that a narrow cone of rays
emitted about a field line with initially no average polarization will
eventually become nearly completely polarized and that the electric
fields will lie in the plane of the dipole field line, if the cone
and the direction of the magnetic field become sufficiently
misaligned.  For radiation due to longitudinal acceleration parallel
to $\bf B$, the opening half angle of the cone is of order $\gp^{-1}
\approx 0.01~\rm{rad}$.  The angle between the cone and the field line
at the transition point is of order $\theta_f \approx 0.1~\rm{rad}$;
hence, our assumption seems reasonable.

\section{Results From Applying The Model To The Vela Pulsar}

Our model has eight parameters.  Six of these describe the pulsar:
$R_\ast$ is its radial size; $T_\ast$, its rotation period; $L_\ast$,
its spin-down luminosity; $\iota$, the angle between the magnetic
field axis and the rotation axis; $\gp$, the bulk relativistic factor
of the electron-positron wind; and $\sigt$, the width of the beam.
The two remaining parameters $\alpha$, the angle between the rotation
axis and the line of sight, and $\omega$, the observing frequency,
describe the observer.

All of the parameters are set by observations, theory, or the observer
except for $\alpha$, $\iota$, $\sigt$, and $\gp$.  However,
observations of polarization often determine the quantity
$\alpha-\iota$, and sometimes determine $\iota$ as well.  Thus, our
model contains, in effect, two or three parameters that can be set to
change the pulse shape.

The degree to which refractive effects dominate the propagation of the
electromagnetic waves depends mainly upon the plasma density on the
open field lines.  We calculated this density by assuming that the
spin-down luminosity of the pulsar is carried away by the electrons
and positrons on the open field lines; thus, the ratio $L_\ast/\gp$
sets the density of the electron-positron wind at a given radius.  We
chose to apply the model to the case of the Vela pulsar because its
relatively high spin-down luminosity
$L_\ast=6.918\times10^{36}~\rm{erg~s}^{-1}$ and correspondingly dense
plasma should emphasize the refractive effects.

Vela's short $89.30~\rm{ms}$ rotation period translates to a light
cylinder radius of $4236~\rm{km}$.  We found that the transitions from
slow-mode to fast-mode waves occur within 10\% of this distance.
Relativistic effects, small as they are, could still have a
significant effect on the pulse shape.

We set $\iota$ to $90\arcdeg$ and $R_\ast$ to $10~\rm{km}$ and chose
the values for the remaining parameters $\gp=100$,
$\sigt=0.020~\rm{rad}$, and $\alpha=94.8\arcdeg$ so that the model
would produce a reasonable pulse shape for an observing frequency of
2.3~GHz.  We did not perform an extensive search of the parameter
space to study the effects on the pulse profile.

Figure~\ref{raytrace} depicts the paths of several rays to the
observer as seen in the corotating frame.  In this frame, observers
follow paths which encircles the magnetic pole and see a pulse when
they cuts across the ``spray'' of light.  From this figure, we also
see that observers would measure the size of the pulsar, i.e. the size
of the transition surface, to be around 100 km at 2.3~GHz, a result
which agrees reasonably well with recent size measurements of Vela.
\citet{Mac99} obtained an upper limit of 50 km from 660~MHz
scintillation data and suggested that the size of the emission region
increases with frequency; \citet{Gwi00} measured the size of Vela's
emission region to be 440~km; and \citet{Kri83} found a spread in
altitude of about 400 km.

\begin{figure}
\plotone{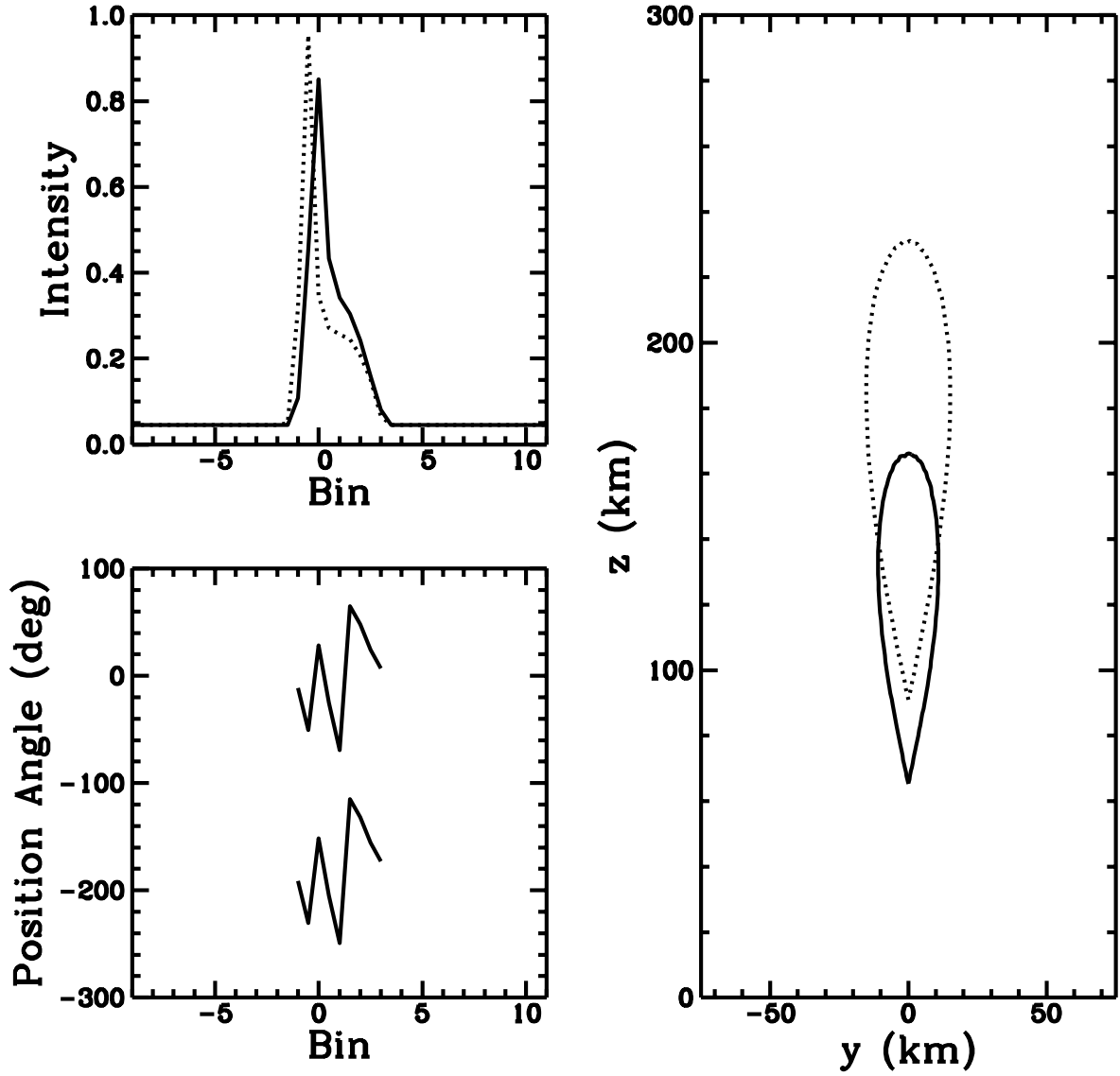}
\caption{This figure shows model-generated pulse profiles at 1.4~GHz
(dotted) and 2.3~GHz (solid) and the corresponding transition surfaces.
The pulse period consists of one hundred bins.}
\label{profile}
\end{figure}

Figure~\ref{profile} shows the intensity profiles obtained from the
model for two frequencies and the associated transition surfaces.  The
dotted curves correspond to 1.4~GHz, and the solid curve, to 2.3~GHz.
The larger transition surface at 1.4~GHz results in a wider pulse.  A
plot of the position angle shows a smooth sweep, somewhat obscured by
the break in the curve caused by the coarse graining of the bins.  At
the start of the pulse, the shape of the curve deviates from the
S-shaped curve predicted by the simple rotating-vector model because
rays from many longitudes contribute to the leading edge of the pulse.

\begin{figure}
\epsscale{0.6}
\plotone{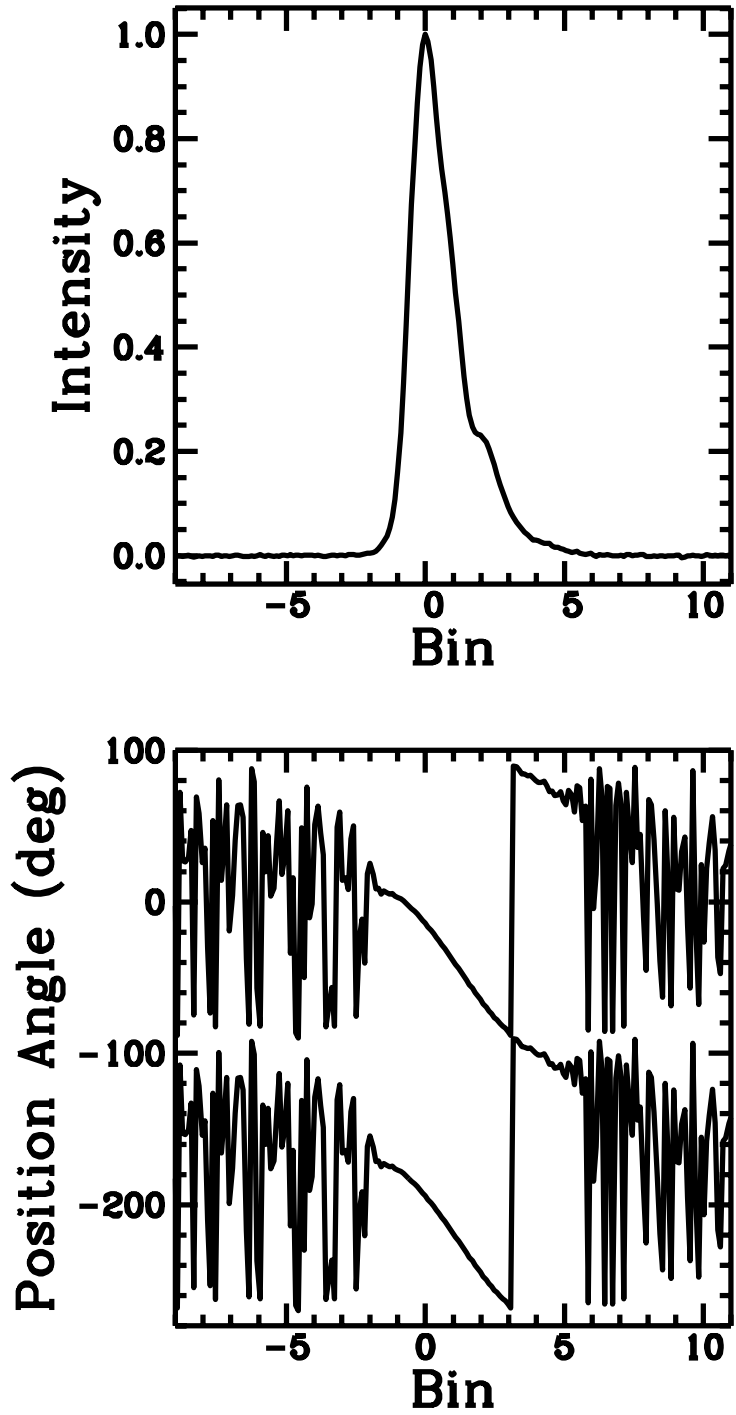}
\caption{The observed integrated pulse profile of the Vela pulsar at
1.4~GHz.  We thank S. Sallmen and D. C. Backer for providing these
data.}
\label{vela}
\end{figure}

Figure~\ref{vela} shows the observed integrated pulse profile for the
Vela pulsar at 1.4~GHz.  The average position angle sweeps through
roughly the same range for both profiles.  More interesting, the
predicted and observed profiles both appear to consist of the overlap
of a strong subpulse followed by a weaker subpulse, although the model
did not reproduce the relative strengths accurately.  In our model,
the presence of two subpulses arises from a combination of three
effects.

Every ray reaching the observer emanates from a point on the stellar
surface; therefore, we can map a pulse to a curve or a region, if rays
from multiple points reach the observer at the same time, on the
stellar surface.  Moving along this curve corresponds to advancing in
time on the profile.  Typically, this path starts at some point
$(\theta_0,\phi_0)=(\Theta_0,\Phi_0)$ and while $\phi_0$ monotonically
decreases, $\theta_0$ first decreases, reaches a minimum, and then
increases.

Because of the cylindrical symmetry around the magnetic pole, each
circle of latitude on the stellar surface has an associated circle of
transition points in the magnetosphere.  The density of rays reaching
the observer is higher when the line of sight grazes one of these
circles than when it cuts across the circle at some angle.  Hence, the
density of rays reaches a peak when $\theta_0$ hits its minimum and
the path on the stellar surface is tangent to a circle of latitude.

Each ray represents an area element $d\Omega_0\approx\theta_0
d\theta_0 d\phi_0$, so its weight contains a factor of $\theta_0$ as
$d\theta_0$ and $d\phi_0$ are fixed in our model.  The weight assigned
to a ray also has the Gaussian factor mentioned earlier, representing
the strength of an average beam.  The beam is strongest at the
magnetic pole and weakens away from the pole with a characteristic
distance specified by $\sigt$.

This combination of weighting factors can easily lead to the
appearance of two subpulses.  At first, when $\theta_0$ is large, the
Gaussian factor suppresses the weighting. As $\theta_0$ decreases, the
weighting at first steadily increases.  At some point, the area factor
$\theta_0$ will begin to dominate, and the weight will decrease,
resulting in the first subpulse.  After $\theta_0$ reaches its
minimum, the reverse sequence occurs, generating the second subpulse.

Neither the higher density of rays at smaller values of $\theta_0$ nor
the weighting assigned to the rays can result in the observed pulse
with its asymmetry between the two subpulses.  Neither effect nor
their combination breaks the cylindrical symmetry.  If the density
effect enhances one weighting maximum, it will enhance the other one
equally.  With only these two effects, the observer would always see
symmetrical pulse.

The final effect is the time-of-arrival effect, which does break the
symmetry.  In the corotating frame, the cylindrical symmetry about the
magnetic pole results in a transition surface with a reflection
symmetry through the x-z plane.  The Lorentz boost to the observer's
frame, however, breaks the symmetry by rotating all of the wave
vectors in the direction of the boost and, because this rotation
affects the time-of-arrival, skews the pulse to the left in the profile.
The other contribution to the time-of-arrival delay further skews the
pulse to the left; rays from further out on the transition surface
take less time to travel to the observer and arrive earlier in the
pulse. Also because their transition points are moving faster, they
undergo a larger boost, resulting in a larger rotation and causing
them to arrive even earlier in the pulse.

\begin{figure}
\plotone{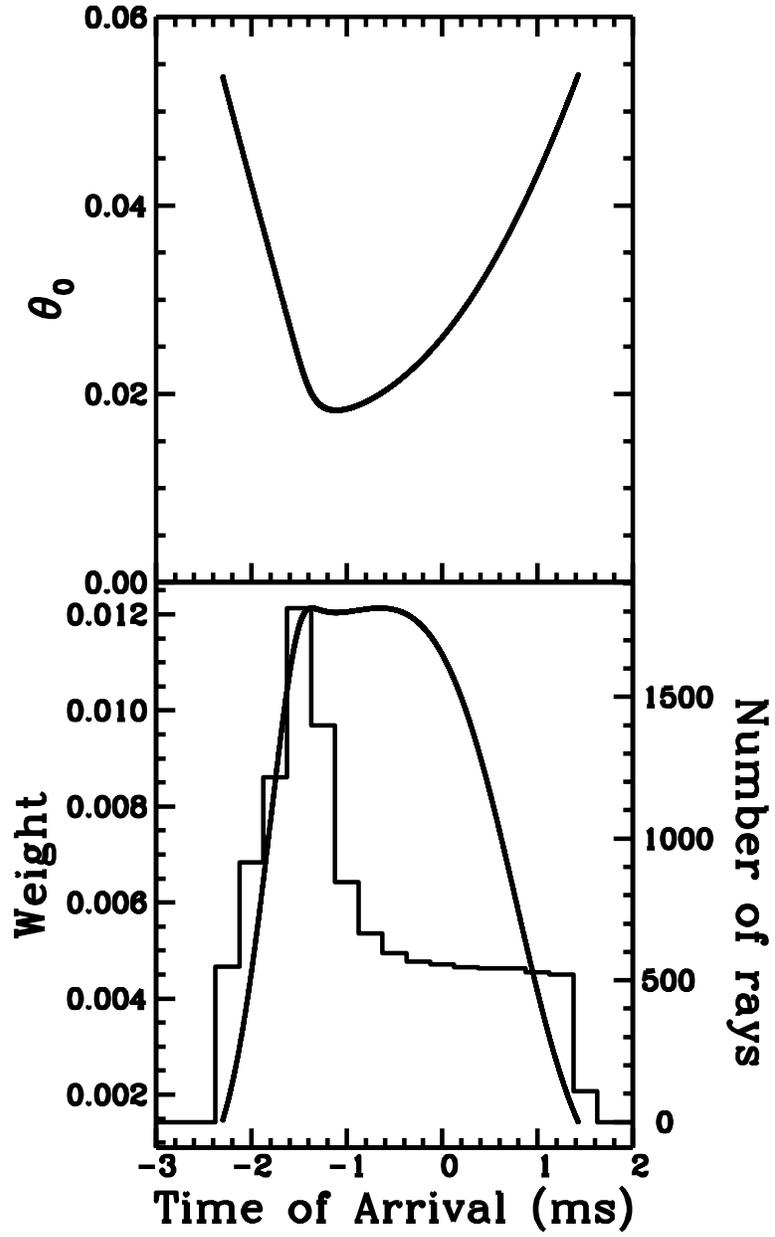}
\caption{Rays from near the magnetic pole arrive early in the pulse,
enhancing the first subpulse.}
\label{anatomy}
\end{figure}

Figure~\ref{anatomy} illustrates how these effects combine to give the
observed pulse.  The top plot shows that rays starting near the
magnetic pole arrive early in the pulse.  The bottom plot shows the
weighting with two maxima, but these small variations can not alone
account for the large difference in strength between the two
subpulses.  The superimposed histogram shows that many rays amass
near the beginning of the profile.  The first subpulse results then
from a combination of the first maximum in the weighting and the large
number of rays reaching the observer at the same time.  The second
subpulse arises as the rays arrive at a constant rate but 
with decreasing weights.

\section{Summary}

A simple ray-tracing model that assumes a neutron star with a
corotating dipolar magnetic field yields results which appear
qualitatively consistent with observations.  The resulting pulse
profile exhibits the general characteristics of the observed profiles;
the size of the emission region agrees with recent measurements; and
the model naturally incorporates the rotating-vector model if one
assumes the mode conversion process does not obliterate polarization
information.

The basic nature of the model, moreover, leaves room for refinement.
We have neglected the physics in the transition from the slow mode to
the fast mode, and an improved model may involve a more realistic
momentum distribution for the plasma.  Finally, adjusting the weight
assigned to each ray by incorporating a more detailed model of the
radiation distribution at the surface may yield information about the
shape of individual pulses and their variations.

\appendix
\section{Coordinate Transformations From the Corotating Frame to the Observer's Frame}

After solving equation~(\ref{rho}) for $r/{r_0}$, we can find $\theta_f$ using
equation~(\ref{dpf}).
Due to the cylindrical symmetry of the dipole field, transitions occur for all values of $\phi$.
For a given value of $\phi$, the transition occurs at the point
\begin{equation}
{\bf r} = r (\sin{\theta_f}\cos{\phi},\sin{\theta_f}\sin{\phi},\cos{\theta_f}),
\end{equation}
and the ray has a wave vector
\begin{equation}
{\bf k} = \omega (\sin{\xi}\cos{\phi},\sin{\xi}\sin{\phi},\cos{\xi})
\end{equation}
where $\xi = (9/8)\theta_f$.  The electric field $\bf e$
and magnetic field $\bf b$ are given by
\begin{mathletters}
\begin{eqnarray}
{\bf e} & = & (\cos{\xi}\cos{\phi},\cos{\xi}\sin{\phi},-\sin{\xi})           \\
{\bf b} & = & (-\sin{\phi},\cos{\phi},0).
\end{eqnarray}
\end{mathletters}
We are working in units where $c=1$. The vectors are normalized as we are not concerned with their magnitudes but only
with their directions so that we can determine the polarization of the ray as seen
by the observer.

The coordinates in the corotating frame and the observer's frame are related
via two rotations so that
\begin{mathletters}
\begin{eqnarray}
{\bf R} & = & (X,Y,Z) = R_Z(\delta) R_Y(\iota) {\bf r} \\
{\bf \tilde{K}} & = & (\tilde{K}_X,\tilde{K}_Y,\tilde{K}_Z) = R_Z(\delta) R_Y(\iota) {\bf k} \\
{\bf \tilde{E}} & = & (\tilde{E}_X,\tilde{E}_Y,\tilde{E}_Z) = R_Z(\delta) R_Y(\iota) {\bf e} \\
{\bf \tilde{B}} & = & (\tilde{B}_X,\tilde{B}_Y,\tilde{B}_Z) = R_Z(\delta) R_Y(\iota) {\bf b}
\end{eqnarray}
\end{mathletters}
where $R_Z$ and $R_Y$ are the rotation matrices for rotations by the
specified angles about the $Z$-axis and $Y$-axis respectively.

\section{Lorentz Transformations}

The vector $\bf \tilde{K}$ does not give the wave vector of the ray as seen by
the observer; rather, it is the representation of the wave vector as measured
in the corotating frame expressed in the coordinates of the observer's frame.
To find the wave vector as seen by the observer, we must apply a Lorentz
transformation because the transition point moves with a velocity $\bbeta =
(\Omega {\bf \hat{Z}})\times{\bf R}/c$.

The components of $\bf \tilde{K}$ parallel and perpendicular to $\bbeta$ are
given by
\begin{mathletters}
\begin{eqnarray}
{\bf \tilde{K}}_\shortparallel & = & ({\bf \tilde{K}} \cdot \hat{\bbeta}) \hat{\bbeta} \\
{\bf \tilde{K}}_\perp & = & {\bf \tilde{K}} - {\bf \tilde{K}}_\shortparallel
\end{eqnarray}
\end{mathletters}
where $\hat{\bbeta}$ denotes the unit vector in the direction of \bbeta.
According to the observer, the ray has a wave vector
${\bf K} = {\bf K}_\shortparallel + {\bf K}_\perp$ 
where
\begin{mathletters}
\begin{eqnarray}
{\bf K}_\shortparallel & = & \gamma (\tilde{K}_\shortparallel + \beta \omega) \hat{\bbeta} \\
{\bf K}_\perp & = & {\bf \tilde{K}}_\perp.
\end{eqnarray}
\end{mathletters}
and $\gamma = 1/\sqrt{1-\beta^2}$ is the usual relativistic factor.

We must likewise transform the electric and magnetic fields.  The electric and magnetic fields
that the observer will see are given by
\begin{mathletters}
\begin{eqnarray}
{\bf E} & = & \gamma({\bf \tilde{E}}-\bbeta\times{\bf \tilde{B}})-\frac{\gamma^2}{\gamma+1}(\bbeta\cdot{\bf \tilde{E}}) \bbeta \\
{\bf B} & = & \gamma({\bf \tilde{B}}+\bbeta\times{\bf \tilde{E}})-\frac{\gamma^2}{\gamma+1}(\bbeta\cdot{\bf \tilde{B}}) \bbeta.
\end{eqnarray}
\end{mathletters}
The position angle $\psi$ is then given by
\begin{equation}
\psi = \arctan{\left(\frac{E_Y}{\sqrt{E^2-E_Y^2}}\right)}
\end{equation}

%
%

\clearpage

\end{document}